\documentclass[12pt]{article}
\usepackage[cp1251]{inputenc}
\usepackage[russian]{babel}

\title{Алгебра Пуанкаре в гамильтоновом формализме релятивистской теории гравитации}
\author{В.О. Соловьев, М.В. Чичикина}
\begin{document}
\maketitle
\begin{abstract}
Показано, что соответствующим образом выбирая входящие в
гамильтониан релятивистской теории гравитации (РТГ) произвольные
функции, можно получить генераторы группы Пуанкаре. Их скобки
Дирака реализуют алгебру группы, в согласии с тем, что в РТГ
имеются 10 интегралов движения.
\end{abstract}

\section{Введение}
Релятивистская теория гравитации~\cite{Log} (РТГ), как почти все теории поля, может быть представлена в общековариантном виде. Пространство-время Минковского является в ней нединамической структурой, и РТГ инвариантна относительно группы Пуанкаре. В этой работе мы продемонстрируем особенности канонической реализации группы Пуанкаре в РТГ, близко следуя подходу Дирака~\cite{Dirac}, а также Редже и Тейтельбойма~\cite{RT}. Мы начинаем с вычисления скобки Пуассона (которая в данном случае заменяется скобкой Дирака) двух гамильтонианов вида
\begin{equation}
H(N,N^i)=\int\left(N{\cal H}+N^i{\cal H}_i\right)d^3x,\label{eq:alg}
\end{equation}
различающихся входящими в них произвольными функциями
пространственных координат $N(x)$, $N^i(x)$. Согласно Дираку (см.
также работу Тейтельбойма~\cite{Teit}), для общековариантных
теорий поля должны выполняться соотношения вида
\begin{equation}
\{H(\alpha, \alpha^i),H(\beta,\beta^j)\}=H(\lambda, \lambda^k),
\label{eq:alg1}
\end{equation}
где
\begin{eqnarray}
\lambda&=&\alpha^i\beta_{,i}-\beta^i\alpha_{,i},\nonumber\\
\lambda^k&=&\gamma^{k\ell}(\alpha\beta_{,\ell}-\beta\alpha_{,\ell})+\alpha^\ell\beta^k_{,\ell}-\beta^\ell\alpha^k_{,\ell}.\label{eq:algebra_par}
\end{eqnarray}
Поскольку в эти соотношения входят не только параметры $\alpha$,
$\alpha^i$, $\beta$, $\beta^j$, но и контравариантные компоненты
метрики $\gamma^{k\ell}$, Тейтельбойм показал, что в ОТО, где
имеется единственная метрика пространства-времени, и она является
динамической переменной, величины ${\cal H}$, ${\cal H}_i$ должны
быть связями.

При наличии фонового пространства, и соответственно,
нединамической метрики в гамильтониане, требуется, чтобы она также
изменялась, как при преобразованиях координат на гиперповерхности,
так и при деформациях гиперповерхности. Это обычно достигается
введением дополнительных гамильтоновых переменных: функций
вложения $e^\alpha(t, x^i)$ и сопряженных к ним импульсов
$p_\alpha(t, x^i)$. Мы не будем вводить этих переменных, и поэтому
наша скобка для гамильтонианов, зависящих от произвольных функций,
не будет полностью совпадать с ответом Дирака и Тейтельбойма. Тем
не менее, мы считаем полезным привести явный ответ для этой
скобки. Переход к узкому классу произвольных функций,
соответствующему алгебре Пуанкаре, будет проводиться на базе
полученного общего результата.

\section{Гамильтонов формализм РТГ}

За динамические переменные в РТГ могут быть взяты, например, 10
компонент римановой метрики и поля материи (здесь это будет
массивное  скалярное поле без самодействия). Соответствующие 10
компонет плоской метрики Минковского считаются известными
функциями координат пространства-времени. Состояние задается на
каждой из однопараметрического (параметр $t$ считается временем)
семейства пространственноподобных гиперповерхностей, которая в
свою очередь задается четырьмя функциями пространственных
(внутренних) координат $X^\alpha=e^\alpha(t,x^i)$ и из метрики
выделяются 6 компонент индуцированной метрики
$\gamma_{ij}=g_{\mu\nu}e^\mu_i e^\nu_j$ (здесь сделано
традиционное~\cite{Kuchar} упрощение обозначений:
$e^\alpha_i\equiv e^\alpha_{,i}$), точно так же, как в ОТО.
Биметризм проявляется в возможности по-разному записать
гамильтониан РТГ:
\begin{eqnarray}
H&=&\int\left(N{\cal H}+N^i{\cal H}_i\right)d^3x\label{eq:Ham2}\\
&\equiv& \int\Biggl(\bar N\bar{\cal H}+\bar N^i\bar{\cal H}_i\nonumber\\
&+&\frac{m^2\sqrt{\eta}}{\kappa}N\left[
-1-\frac{f^{\perp\perp}}{2}+
\frac{f^{\perp i}f^{\perp j}\eta_{ij}}{2f^{\perp\perp}}-
\frac{1}{f^{\perp\perp}}\frac{\gamma}{\eta}\left(\frac{1}{2}
\eta_{ij}\gamma^{ij}-1
\right)
\right]\Biggr),\label{eq:Ham3}
\end{eqnarray}
где
\begin{eqnarray}
\bar N&=&-\frac{1}{f^{\perp\perp}}\sqrt{\frac{\gamma}{\eta}}N,\quad
\bar N^i=N^i-\frac{f^{\perp i}}{f^{\perp\perp}}N,\nonumber\\
{\cal H}&=&-\frac{1}{f^{\perp\perp}}\sqrt{\frac{\gamma}{\eta}}{\bar{\cal H}}-
\frac{f^{\perp i}}{f^{\perp\perp}}{\bar{\cal H}}_i\nonumber\\
&+&\frac{m^2\sqrt{\eta}}{\kappa}\left[
-1-\frac{f^{\perp\perp}}{2}+
\frac{f^{\perp i}f^{\perp j}\eta_{ij}}{2f^{\perp\perp}}-
\frac{1}{f^{\perp\perp}}\frac{\gamma}{\eta}\left(\frac{1}{2}
\eta_{ij}\gamma^{ij}-1
\right)
\right],\nonumber\\
\bar{\cal H}&=&\frac{1}{\sqrt{\gamma}}\left(-
\frac{1}{\kappa}\gamma\tilde R+\kappa(\mathrm{Sp}\pi^2
-\frac{\pi^2}{2}) +
\frac{\pi^2_{\phi}}{2}+\frac{1}{2}\gamma\gamma^{ij}\partial_i\phi\partial_j\phi
+\frac{1}{2}\gamma M^2\phi^2
\right),\nonumber\\
{\cal H}_i&=&\bar{\cal H}_i=-2\pi_{i|j}^j+\pi_{\phi}\phi_{,i}.
\end{eqnarray}
Черта над величиной обозначает определение этой величины на основе
римановой метрики, а соответствующие величины без черты
определяются на основе метрики Минковского. Как было показано в
работе~\cite{SoChi}, 4 проекции  тензора римановой метрики вдоль
нормали и сопряженные к ним импульсы могут быть исключены из числа
независимых гамильтоновых переменных разрешением связей 2-го рода:
\begin{eqnarray}
f^{\perp i}&=&\frac{\kappa}{m^2\sqrt{\eta}}\eta^{ij}\bar{\cal H}_j,
\label{eq:f_i}\\
f^{\perp\perp}&=&-\frac{\kappa}{m^2\sqrt{\eta}}\sqrt{
\eta^{ij}\bar{\cal H}_i\bar{\cal H}_j+2\frac{m^2\sqrt{\eta}}{\kappa}
\left[
\sqrt{\frac{\gamma}{\eta}}\bar{\cal H}+\frac{m^2\sqrt{\eta}}{\kappa}
\frac{\gamma}{\eta}\left(\frac{1}{2}\eta_{ij}\gamma^{ij}-1\right)
\right]}.\nonumber\\
\label{eq:f_perp}
\end{eqnarray}
и переходом к скобкам Дирака, которые в данном случае определяются формулой
\begin{equation}
\{F,G\}=\int\limits_{R^3}d^3x\left[\frac{\delta F}{\delta\gamma_{ij}}
\frac{\delta G}{\delta\pi^{ij}}+\frac{\delta F}{\delta\phi}
\frac{\delta G}{\delta\pi_\phi}
-\frac{\delta F}{\delta\pi^{ij}}\frac{\delta G}{\delta\gamma_{ij}}-\frac{\delta F}{\delta\pi_\phi}
\frac{\delta G}{\delta\phi}
\right].\label{eq:DB}
\end{equation}

При нахождении вариационных производных по переменным $\gamma_{ij}$,
$\pi^{ij}$ гамильтониан берется в виде (\ref{eq:Ham3}), в каком он
был первоначально получен  в
работе~\cite{SoChi}. Тогда
\begin{eqnarray}
\frac{\delta H}{\delta\phi} &=& -\left(\bar N\sqrt{\gamma}\gamma^{ij}\partial_j\phi\right)_{,i}+\bar
N\sqrt{\gamma}M^2
\phi-\left(\bar N^i\pi_\phi\right)_{,i},\\
\frac{\delta H}{\delta\pi_\phi}
&=&
\bar N\frac{\pi_\phi}{\sqrt{\gamma}}+{\bar N}^i\phi_{,i},\\
\frac{\delta H}{\delta\gamma_{ij}}&=&
\frac{1}{2}\bar
N\sqrt{\gamma}(\gamma^{ij}\gamma^{mn}-\gamma^{im}
\gamma^{jn})\partial_m\phi\partial_n\phi+\frac{1}{2}\bar
N\sqrt{\gamma}
\gamma^{ij}M^2\phi^2\nonumber\\
&+&\frac{1}{\kappa}\bar N\sqrt{\gamma}(R^{ij}-\gamma^{ij}R)-
\kappa\frac{\bar
N}{\sqrt{\gamma}}(\pi\pi^{ij}-2\pi^{ik}\pi^j_k)\nonumber\\
&-&\frac{1}{\kappa} \sqrt{\gamma}({\bar
N}^{|ij}-\gamma^{ij}N^{|k}_{|k})-(\pi^{ij}{\bar N}^k)_{|k}+
\pi^{ik}{\bar N}^j_{|k}+\pi^{kj}{\bar N}^i_{|k}\nonumber\\
&-&\frac{m^2}{\kappa}\bar N\sqrt{\gamma}\left[
\gamma^{ij}+\frac{1}{2}\eta_{kl}\left(
\gamma^{ki}\gamma^{lj}-\gamma^{ij}\gamma^{kl} \right) \right],\\
\frac{\delta H}{\delta\pi^{ij}}&=&{\bar N}_{i|j}+{\bar N}_{j|i}+\kappa\frac{2\bar
N}{\sqrt{\gamma}}(\pi_{ij}
-\gamma_{ij}\frac{\pi}{2}).\\
\end{eqnarray}

\section{Генераторы алгебры Пуанкаре}
Среди вариантов гамильтоновой эволюции, разнообразие которых
проистекает из произвола в выборе функций $N(x)$, $N^i(x)$ в
(\ref{eq:Ham2}), содержатся преобразования, сохраняющие метрику
Минковского. В принципе, при их рассмотрении можно было бы не
ограничивать ни класс пространственноподобных гиперповерхностей,
ни системы координат на них, требуя только, чтобы векторные поля
$N^\alpha=\partial X^\alpha/\partial t$ были полями векторов
Киллинга метрики Минковского
\begin{equation}
N_{\alpha;\beta}+N_{\beta;\alpha}=0.
\end{equation}
Тогда $3+1$-разложение этих уравнений привело бы нас к формулам,
содержащим тензор внешней кривизны гиперповерхности (по отношению
к плоской метрике). Если же ограничиться гиперплоскостями, то
формулы упрощаются до следующего вида
\begin{equation}
N_{|ij}=0,\quad N_{i|j}+N_{j|i}=0,\label{eq:parameters}
\end{equation}
а если еще и координаты выбрать декартовыми, то метрика,
индуцированная на гиперплоскостях метрикой Минковского будет
простейшей
$\eta_{ij}\equiv\eta_{\alpha\beta}e^\alpha_ie^\beta_j=\delta_{ij}$
, а ковариантные производные, согласованные с ней, станут обычными
частными производными. Анализ этого случая вполне достаточен для
иллюстрации роли алгебры Пуанкаре.  Мы получаем на гиперплоскостях
функции преобразований~(\ref{eq:parameters})  в виде
\begin{equation}
N=A_kx^k+a,\quad N^i=A_{ik}x^k+a^i,\label{eq:Poincare}
\end{equation}
где
\[A_{ik}=-A_{ki}.\]
Тогда гамильтониан~(\ref{eq:Ham2}), ввиду его линейности по функциям $N(x)$, $N^i(x)$, примет вид
\begin{equation}
H=P^0a-P^ia^i+M^kA_k+\frac{1}{2}M^{ik}A_{ik},\label{eq:Ham_P}
\end{equation}
где
\begin{eqnarray}
P^0 &=&-\frac{m^2}{\kappa}\int \left(1+f^{\perp\perp}\right)d^3x, \\
P_i &=& -\frac{m^2}{\kappa}\int f^{\perp i}d^3x\equiv-\int{\cal H}_i d^3x, \\
M^{ik} &=& -\frac{m^2}{\kappa} \int\left(x^i f^{\perp k}-x^k f^{\perp i}\right)d^3x\equiv\int\left(x^k{\cal H}_i-x^i{\cal H}_k\right)d^3x, \\
M^k &=& -\frac{m^2}{\kappa}\int x^k(1+f^{\perp\perp})d^3x.
\end{eqnarray}
Обозначения формул (\ref{eq:Poincare}), (\ref{eq:Ham_P}) выбраны для удобства сравнения с аналогичными формулами
работы~\cite{RT}, где рассматривалась алгебра Пуанкаре в асимптотически плоском пространстве ОТО.

\section{Алгебра Пуанкаре}
Полезно получить алгебру скобок Дирака для гамильтонианов общего вида. С вычислительной точки зрения, при выводе
удобно использовать  гамильтониан в виде (\ref{eq:Ham3}), а связи 2-го рода учитывать только после вычисления
скобок, что позволяет при вычислениях не принимать во внимание зависимость $\bar N$, $\bar N^i$ от $f^{\perp\perp}$, $f^{\perp i}$. Как обычно, при расчетах отбрасываются все поверхностные интегралы, что оправдано характером
асимптотического поведения метрики в массивной гравитации.

Результаты вычислений представим в форме, удобной для сравнения с аналогичной формулой ОТО:
\begin{eqnarray}
\{H(\alpha, \alpha^i),H(\beta,\beta^j)\}&=&\int d^3x \Biggl[\bar\lambda\bar{\cal H}+
\bar\lambda^k\bar{\cal H}_k\nonumber\\
&+& \left(\bar\alpha(\bar\beta^{k|\ell}+\bar\beta^{\ell
|k})-\bar\beta(\bar\alpha^{k|\ell}+\bar\alpha^{\ell|k})\right)\nonumber\\
&\times& \left(\frac{1}{2}\gamma_{k\ell}\bar{\cal H}
-\frac{m^2}{\kappa}\sqrt{\gamma}\gamma_{k\ell}
(1-\frac{1}{2}\eta_{mn}\gamma^{mn})-\frac{m^2}{\kappa}\frac{\sqrt{\gamma}}{2}\eta_{k\ell}\right)
\Biggr],\nonumber\\ 
\label{eq:algebra}
\end{eqnarray}
где
\begin{eqnarray}
\bar\lambda&=&\bar\alpha^i\bar\beta_{,i}-\bar\beta^i\bar\alpha_{,i},\nonumber\\
\bar\lambda^k&=&\gamma^{k\ell}(\bar\alpha\bar\beta_{,\ell}-\bar\beta\bar\alpha_{,\ell})+\bar\alpha^\ell\bar\beta^k_{,\ell}-\bar\beta^\ell\bar\alpha^k_{,\ell}.
\end{eqnarray}
Отличия от ОТО проявляются здесь как в появлении членов,
пропорциональных квадрату массы гравитона, так и в коэффициенте
при $\bar{\cal H}$. Последнее связано с тем, что коэффициент при
$\bar{\cal H}$, т.е. функция $\bar N$, пропорционален
$\sqrt{\gamma}$. Соотношения (\ref{eq:algebra}), таким образом, не
представляют собой алгебру деформаций гиперповерхности
(\ref{eq:alg1}), (\ref{eq:algebra_par}), т.к. функции $\bar N$,
$\bar N^i$ не являются ее параметрами и т.к. мы не включили в
скобку переменные $e^\alpha$, $p_\alpha$.

Подстановка в соотношения~(\ref{eq:algebra}) вместо произвольных функций $\alpha$, $\alpha^i$,
$\beta$, $\beta^j$ выражений вида (\ref{eq:Poincare}), отвечающих
преобразованиям Пуанкаре, приводит к соотношениям алгебры Пуанкаре:
\begin{eqnarray}
  \{P^0,P_i\} &=& 0, \quad  \{P_i,P_j\} = 0, \\
  \{P^0,M^{ik}\} &=& 0, \quad  \{P_i,M^{jk}\} = \delta_{ik}P_j-\delta_{ij}P_k, \\
  \{M^{ij},M^{k\ell}\}
&=&\delta_{ik}M^{j\ell}-\delta_{i\ell}M^{jk}+\delta_{j\ell}M^{ik}-\delta_{jk}M^{i\ell}, \\
  \{P^0,M^i\} &=& -P^i, \quad  \{P_i,M^j\} =  -\delta_{ij}(P^0-c^0),\\
 \{M^k,M^{ij}\}&=&\delta_{kj}(M^i-c^i)-\delta_{ki}(M^j-c^j),\quad \{M^i,M^j\}=-M^{ij}.
\end{eqnarray}
Аддитивные вклады $c^0=m^2/\kappa\int d^3x$ и $c^i=m^2/\kappa\int x^id^3x$ в $P_0$ и $M^i$, не зависящие от канонических
переменных и выражаюшиеся расходящимися интегралами по всему
пространству, играют роль центральных зарядов в канонической
реализации алгебры Пуанкаре и отвечают классической перенормировке
энергии вакуума.

\section{Заключение}

С одной стороны, РТГ является теорией поля на фиксированном плоском фоне и поэтому к ней должны быть применимы результаты Дирака относительно параметризованных теорий поля.

С другой стороны, язык РТГ близок языку ОТО и в качестве гамильтоновых переменных в ней могут быть выбраны те же самые величины $\gamma_{ij}$   и $\pi^{ij}$.

В ОТО мы получаем алгебру деформаций гиперповерхности без
включения в число переменных $e^\alpha$ и $p_\alpha$. В
параметризованных теориях поля это включение необходимо.  Мы
останавливаемся в промежуточной позиции -- используем
$\gamma_{ij}$, $\pi^{ij}$ и метрику Минковского без  $e^\alpha$,
$p_\alpha$  -- и соответственно, получаем соотношения, отличные от
(\ref{eq:alg1}). Эти соотношения достаточны  для перехода от них к
алгебре Пуанкаре.

Различия между РТГ и ОТО здесь весьма существенны.

Во-первых, в ОТО величины $\bar{\cal H}$ и $\bar{\cal H}_i$ являются связями, т.е. обращаются в нуль на решениях уравнений движения. В РТГ эти величины отличны от нуля и например  $\bar{\cal H}_i$ оказывается пропорциональной плотности импульса.

Во-вторых, в ОТО поля (величины $\gamma_{ij}-\eta_{ij}$) убывают медленно, и при вычислении скобок Пуассона, соответствующих алгебре Пуанкаре, необходимо сохранять поверхностные интегралы. В РТГ  все поверхностные интегралы отбрасываются, т.к. $\gamma_{ij}-\eta_{ij}$ убывают быстро.

Отсутствие связей первого рода приводит к тому, что гамильтонианы
РТГ, в частности, генераторы группы Пуанкаре, в отличие от
гамильтонианов ОТО, не сводятся к поверхностным интегралам на
решениях уравнений движения.  Таким образом, в рамках РТГ
определена плотность энергии-импульса и других интегралов
движения.

Вопросы канонического квантования мы надеемся рассмотреть в следующих работах.

\end{document}